\documentclass[12pt,thmsa]{article}

\begin{document}

\title{Fluctuation depinning of the dislocation kinks under the influence of
external forces}
\author{A. S. Vardanyan, A. A. Kteyan, R. A. Vardanyan \\
{\small \textsl{Solid State Laboratory, Institute of Radiophysics {\&}
Electronics,}} \\
{\small \textsl{Armenian National Academy of Sciences, Ashtarak-2, 0203,
Armenia}}\\}
\maketitle

\begin{abstract}
Within the framework of McLaughlin-Scott perturbation theory the
equation-of-motion of the dislocation kink in the pinning potential is
linearized, assuming the simultaneous influence of \textit{ac} and \textit{dc%
} forces. Based on the equations derived, the probability of kink depinning
was studied. The dependencies of the depinning probability on \textit{dc}
and \textit{ac} forces are analyzed in details.
\end{abstract}



\section{Introduction}

In many elemental and compound semiconductors the motion of dislocations is
controlled by Peierls mechanism, i.e. it occurs through formation and
migration of kink and anti-kink pairs. Formation of the dislocation kink
implies that various dislocation segments are located in the neighboring
valleys of the periodic relief (Peierls potential) of the lattice. Motion of
a dislocation in Peierls relief is typically divided into three phases \cite
{Suz},\cite{Ma}: formation of kink pairs, migration of the kink and
anti-kink in opposite directions and annihilation of kinks with different
signs moving towards each other.

It is well known, that various point defects (impurity atoms, vacancies) may
impede the motion of the kink along the dislocation, i.e. they can act as
pinning centers for the kink. In absence of lattice strain, the kink can be
held in the potential of the pinning centre, while the influence of a
stationary force can detach it with certain probability. Simultaneous effect
of \textit{ac} force can change (increase or decrease) the depinning
probability. This effect was theoretically analyzed for the case, when a
segment of the dislocation line overcomes the pinning centre \cite{Va}.

The importance of the studied problem is mostly due to the issue of
recombination - enhanced motion of dislocations. In semiconductors the
electronic features of the dislocation kink can influence the dislocation
dynamics, provided that the multiphonon capture of the charge carriers by
the kink is the significant mechanism of their recombination. During the
multiphonon capture the lattice fluctuations localized in proximity to the
capture centre are stimulated. Subsequently, those fluctuations are
dispersed through the entire crystal. Thus, if recombination of charge
carriers involves the dislocation kinks and happens through the thermal
capture mechanism \cite{VVK}, then the released energy excites additional
fluctuations which may stimulate the motion of dislocation (similarly the
photoplastic effect in semiconductors has been interpreted \cite{VKO}). It
is generally accepted that for dislocation motion by the Peierls mechanism,
recombination basically contributes to the kink migration along the
dislocation \cite{Ma},\cite{Pir}. Recombination - enhanced motion of
dislocations leads to extension of stacking faults in junction regions of
bipolar and optoelectronic semiconductor devices and eventually results in
the device degradation.

Thus, fluctuation-assisted depinning of the dislocation kink represents a
single act that causes degradation of a semiconductor device. Therefore, the
aim of our work is exploring stochastic dynamics of the pinned kink under
the influence of stationary and alternating (harmonic) forces. During the
degradation process the dislocation is influenced by constant force stemming
from internal tensions of the lattice, and the lattice fluctuations
resulting from thermal recombination of carriers act as the \textit{ac}
force.

Kinks and anti-kinks on the dislocation are non-linear objects (described by
the sin-Gordon (SG) soliton equation). Therefore, the study of
fluctuation-assisted depinning of the kink is a complicated task. Simple
approach to solving this problem is used in Ref. \cite{Zak}, where dynamics
of the kink is described by the equation of linear harmonic oscillator,
which has the kink mass ascribed to it. This simplification allows to
basically describe the process, however, for the purpose of further
development of theory (to fully account for the dependence of detachment
probability on the kink parameters), it is necessary to correctly linearize
the equation of motion.

Linearization of the equation of motion of the kink is possible through the
use of perturbation theory developed by McLaughlin and Scott \cite{Mc}.
Although this theory is based on some assumptions, it is actually the only
reliable approach used in many studies for exploring the effect of external
forces on the kink (see Ref. \cite{San} and references therein). In this
paper, we will use a modification of above-mentioned perturbation theory
developed in \cite{Gul}. When linearizing the equations of the kink motion
in presence of \textit{ac} force, this method enables to avoid complicated
calculations used by the authors \cite{Mc}.

\section{Kink dynamics under influence of ac\ force}

Let's consider a smooth kink on a dislocation (directed along $x$ axis),
i.e. the width $w$ significantly exceeds the height $a$ (Fig.1). The height $%
a$ of the kink (equal to the inter-atomic distance) represents a period of
Peierls relief. The kink is described by the transversal displacement $%
\varphi $ of the dislocation, which is equal to zero when $x\to -\infty $,
and equal to $a$ when $x\to \infty $ (for anti-kink the reverse is true,
i.e. $\varphi $ decreases from $a$ to zero). The kink width corresponds to
the range where $0<\varphi <a$. The smallness of the parameter $\tau _P/G$
is the criterion determining the smoothness of the kink, where $\tau _P$ is
the Peierls tension and $G$ is the shift module. The kink width relates to
its height as per following equation \cite{Suz}:

\begin{equation}
w=a\left( \frac{\pi /2}{\tau _P/G}\right) ^{1/2}.  \label{eq1}
\end{equation}
Therefore, the inequality ${\tau }_P{/}G<<1$ denotes that the smooth kinks
are being formed, as from (\ref{eq1}) we get the following: $w>>a$. In this
case the string model of the dislocation is applicable for description of
the kink.

If the kink is located close to the pinning point, we can describe the
interaction of the kink with the pinning centre (located at the point $x=0$
on the dislocation) by the following potential \cite{Mc},\cite{Gul}: 
\begin{equation}
U=-\frac a{2\pi }\mu \delta (x)\left( {1-\cos }\frac{2\pi \varphi }a\right) ,
\label{eq2}
\end{equation}
where parameter $\mu $ characterizes the interaction force. The multiplier $%
1-\cos \left( 2\pi \varphi /a\right) $ in equation (\ref{eq2}) ensures that
the pinning potential $U$ becomes equal to zero outside of the kink region.

As indicated in the Introduction, we are interested in the kink dynamics
under influence of \textit{dc} and \textit{ac} forces. The constant force $%
\Gamma $ stemming either from the internal pressure or the lattice external
strain tends to move the kink away from the pinning centre. We will consider
the \textit{ac} force $\gamma (t)$, which emulates the lattice fluctuations
originating due to the recombination process, as harmonic process. Assuming
that the energy of these forces, as well as dissipation due to friction are
small compared to Peierls potential, we can consider the influence of
external forces as small perturbation relative to the undisturbed system. If
the periodic potential has the following form (commonly accepted for Peierls
relief \cite{Suz},\cite{HL}): 
\begin{equation}
V_P=\frac{a^2\tau _P}{2\pi }\sin ^2\left( \frac{\pi \varphi }a\right) ,
\label{eq3}
\end{equation}
then the dynamics of a string in this potential in case of existence of
dissipation can be described by disturbed SG equation \cite{Mc}: 
\begin{equation}
m\varphi _{tt}-\chi \varphi _{xx}=-\frac{\pi a\chi }{2w^2}\sin \frac{2\pi
\varphi }a+f,  \label{eq4}
\end{equation}
where $m$ and $\chi $ are the mass and the linear tension of the dislocation
(per unit length) respectively. The force $f$ is the sum of all acting
forces: 
\begin{equation}
f=-\eta \varphi _t-\Gamma -\gamma (t)+\mu \delta (x)\sin \frac{2\pi \varphi }%
a,  \label{eq5}
\end{equation}
where $\eta $ characterizes dislocation damping per unit length ($\eta >0)$.
For derivation of equation (\ref{eq4}) it was taken into account that $\chi
\simeq {Gb^2/}2$ \cite{HL}, where $b$ is the module of the Burgers vector
(which has the order of $a$), $s$ is the sound velocity in solids, and the ${%
\tau }_P{/}G$ ratio is determined by equation (\ref{eq1}).

If the system is not perturbed, i.e. $f=0$, then the solution of equation (%
\ref{eq4}) is the following \cite{Mc},\cite{San},\cite{Gul}: 
\[
\varphi =\frac{2a}\pi \arctan \left[ \exp \left( \pm \frac \pi wg\left(
v\right) \left( x-x_0-vt\right) \right) \right] ,
\]
where $x_0$ is the initial coordinate of the kink, $v$ is the velocity of
motion across the dislocation ($v<s)$, and $g(v)=1/\sqrt{1-\left( v/s\right)
^2}$. Here, the positive sign corresponds to the kink and the negative sign
-- to the anti-kink. This solution corresponds to translation invariance of
the kink along the dislocation. The energy of SG system is presented by the
following Hamiltonian \cite{Mc},\cite{Gul}: 
\begin{equation}
H^{SG}=\int\limits_{-\infty }^\infty \left( \frac{w\tau _P}{2s^2}\dot
\varphi ^2+\frac{w\tau _P}2\varphi _x^2+\frac \pi wV_P\right) dx.
\label{eq6}
\end{equation}
By substituting the solution of unperturbed SG equation into (\ref{eq6}) and
integrating over $x$, we will arrive at the Hamiltonian of the unperturbed
system: 
\begin{equation}
H^{SG}=\frac{2a^2\chi }{w^2}\left( {1-}\left( v/s\right) {^2}\right) ^{-1/2}.
\label{eq7}
\end{equation}
If $f<<a\tau _P/2$, i.e. all forces are small compared to the amplitude of
the Peierls potential, then in order to solve equation (\ref{eq4}) we can
use the perturbation technique developed in publications \cite{Mc},\cite{Gul}%
. We assume that solution of equation (\ref{eq4}) has the form of
unperturbed SG equation, however the perturbations lead to modulation of
velocity, i.e. in this case $v$ is a function of time $t$:

\begin{equation}
\varphi =\frac{2a}\pi \arctan \left[ \exp \left( \pm \frac \pi w\left(
g\left( v\right) x-X\left( t\right) \right) \right) \right]   \label{eq8}
\end{equation}
where $X\left( t\right) =\int\limits_0^tg\left( v\left( t^{\prime }\right)
\right) v\left( t^{\prime }\right) dt^{\prime }$ indicates the location of
the kink centre (assuming that $x_0=0$). In absence of terms describing
dissipation and \textit{ac} force in equation (\ref{eq5}), we can write the
Hamiltonian of the perturbed system the following way: 
\begin{equation}
H(\varphi )=H^{SG}+H^p,  \label{eq9}
\end{equation}
where 
\begin{equation}
H^p=\frac \pi w\int\limits_{-\infty }^\infty \left( \Gamma \varphi -\frac
a{2\pi }\mu \delta \left( x\right) \left( 1-\cos \frac{2\pi \varphi }%
a\right) \right) dx  \label{eq10}
\end{equation}
describes the energy of stationary perturbation. In this case $H(\varphi )$
is stationary in time: $dH/dt=0$. When non-stationary forces are present
(the friction force and \textit{ac} external force), the equation of motion
must be determined by the following equation \cite{Sv}: 
\begin{equation}
\frac{dH}{dt}=-\frac \pi w\int_{-\infty }^\infty \left( {\eta \dot \varphi
^2+\gamma }\left( t\right) \dot \varphi \right) dx.  \label{eq11}
\end{equation}
By inserting equation (\ref{eq9}) into (\ref{eq11}), we arrive at the
equation of the momentum of SG system: 
\begin{equation}
\frac{dH^{SG}}{dv}\dot v=\frac \pi w\int\limits_{-\infty }^\infty f{(\varphi
)}\dot \varphi dx,  \label{eq12}
\end{equation}
where $\varphi =\varphi \left( g\left( v\right) x{,X}\right) $ is the
solution of the perturbed SG equation (\ref{eq4}) that have the time
derivative $\dot \varphi \simeq \varphi _Xg\left( v\right) v$ \cite{Gul}.
The equation (\ref{eq12}), along with the equation of time derivate of $X$ 
\begin{equation}
\dot X=g\left( v\right) v,  \label{eq13}
\end{equation}
describes the motion of the kink. The Hamiltonian $H^{SG}$ in equation (\ref
{eq12}) has the same form as the Hamiltonian of the unperturbed SG system (%
\ref{eq6}). The string displacement $\varphi $ is determined by (\ref{eq8}).
By inserting equation (\ref{eq8}) into (\ref{eq6}) and integrating, we
arrive at the expression for the soliton energy (which substitutes the
stationary Hamiltonian (\ref{eq7})): 
\begin{equation}
H^{SG}\left[ \varphi \left( g\left( v\left( t\right) \right) x,X\left(
t\right) ,v\left( t\right) \right) \right] \simeq \frac{2a^2\chi }{w^2}%
\left( {1-}\left( v\left( t\right) /s\right) {^2}\right) ^{-1/2}.
\label{eq14}
\end{equation}
Now, by inserting the expressions (\ref{eq14}) and (\ref{eq5}) into (\ref
{eq12}) and integrating, we get set of differential equations for variables $%
v\left( t\right) $ and $X\left( t\right) $ 
\begin{equation}
{\ }
\begin{array}{c}
M\frac{dv}{dt}=\pm \left[ \Gamma +\gamma \left( t\right) \right] \left(
g\left( v\right) \right) ^{-3}-\frac{2a\eta }{\pi w}v\left( g\left( v\right)
\right) ^{-2}- \\ 
\frac{2\mu }w\left( g\left( v\right) \right) ^{-2}\ sech^2\left( \frac{\pi X}%
w\right) \tanh \left( \frac{\pi X}w\right)  \\ 
\frac{dX}{dt}=g\left( v\right) v
\end{array}
{,}  \label{eq15}
\end{equation}
where we accounted for $\chi =ms^2$; $M=2am/\pi w$ is the effective mass of
the kink. The set of equations (\ref{eq15}) determines the dynamics of the
isolated kink (i.e. the one which does not interact with other kinks)
conditioned by the general force $f$.

Let's explore the motion of the kink in the vicinity of the pinning point.
Equations (\ref{eq15}) are typically analyzed in the phase plane$\left( v{,X}%
\right) $. The singular point, where $\dot v=0$ and $\dot X=0$, in presence
of \textit{dc} force corresponds to the kink equilibrium position \cite{Mc}.
When \textit{ac} force is also in effect, we have to determine the
equilibrium position considering only stationary force, meanwhile the 
\textit{ac} force causes additional shift from the equilibrium. In other
words, the equilibrium position is determined as the kink coordinate
averaged on the time intervals larger than the period of the \textit{ac}
force. As it follows from equation (\ref{eq15}), the following conditions
correspond to the position of equilibrium: $v=0$ and $X=X_0$, where $X_0$ is
the root of the following equation: 
\begin{equation}
\frac{w\Gamma }{2\mu }-\ sech^2\left( \frac{\pi X_0}w\right) \tanh \left( 
\frac{\pi X_0}w\right) =0.  \label{eq16}
\end{equation}
For the purpose of analytical examination of the kink oscillations around
the point $X_0$, we need to linearize the equation (\ref{eq15}) with respect
to variable $v$, assuming that the motion of the kink is non-relativistic ($%
v<<s)$. In absence of \textit{ac} force, the kink makes damping oscillations
around the pinning point, and eventually reaches the rest \cite{Mc}. If the
kink is also subject to the influence of periodic force $\gamma \left(
t\right) =f_0\sin \Omega t$, where $f_0$ is the amplitude of periodic force
and $\Omega $ is its frequency, then we need to analyze the following
equations:

\begin{equation}
{\ }
\begin{array}{c}
M\frac{dv}{dt}=\pm \left[ \Gamma +f_0\sin \Omega t\right] -\frac{2a\eta }{%
\pi w}v-\frac{2\mu }w\ sech^2\left( \frac{\pi X}w\right) \tanh \left( \frac{%
\pi X}w\right)  \\ 
\frac{dX}{dt}=v
\end{array}
{,}  \label{eq16a}
\end{equation}
Further linearization of the equation of motion of the kink is possible in
case of small oscillations, assuming that the displacement is small compared
to the kink width ($X-X_0<<w$). Then we can expand the last term in the
right-hand side of the first equation (\ref{eq16a}) (describing the
interaction of the kink centre with the pinning point) into Taylor's series
around the equilibrium point $X_0$. Keeping only linear in $X-X_0$ term and
using the second expression (\ref{eq16a}), we obtain the equation of linear
oscillations: 
\begin{equation}
\ddot X+2\lambda \dot X+\omega _0^2\left( X-X_0\right) =\pm \frac{f_0}M\cos
\Omega t,  \label{eq17}
\end{equation}
where the parameter $\lambda =a\eta /\pi wM$ is the effective damping
coefficient, and 
\begin{equation}
\omega _0^2=\frac{2\pi \mu }{w^2M}\left[ \ sech^4\left( \frac{\pi X_0}%
w\right) {-2\,}\ sech{^2}\left( \frac{\pi X_0}w\right) {\tanh {^2}}\left( 
\frac{\pi X_0}w\right) \right]   \label{eq18}
\end{equation}
assumes the role of eigenfrequency of the system.

As it follows from equation (\ref{eq16}), when $\Gamma =0$ the equilibrium
position is $X_0=0$, and from equation (\ref{eq18}) we derive $\omega _0^2=%
\frac{2\pi \mu }{w^2M}$. With increasing constant force, $\omega _0$
reduces; the critical value of force $\Gamma _c={4\mu /}\left( {3\sqrt{3}}%
w\right) $ corresponds to the condition $\omega _0=0$, which is determined
from equations (\ref{eq16}) and (\ref{eq18}). In other words, when $\Gamma
>\Gamma _c$ (then equation (\ref{eq16}) haven't roots), the kink overcomes
the pinning potential and can be considered as a free kink.

Note that the constant force does not enter explicitly into equation (\ref
{eq17}). This force determines the equilibrium position $X_0$. Thus,
equation (\ref{eq17}) describes oscillations of the shifted coordinate of
the kink centre under the influence of the harmonic force. In case of small
viscosity $\omega _0>\lambda $, we immediately arrive at the kink
displacement 
\begin{equation}
X=X_0+s\left( t\right) ,  \label{eq19}
\end{equation}
where the function $s\left( t\right) =B\cos \left( {\Omega t+\delta }\right) 
$ describes harmonic displacements; $B$ is the amplitude of the kink
oscillations and $\delta $ is the initial phase: 
\[
B=\frac{f_0}{M\sqrt{\left( {\omega _0^2-\Omega ^2}\right) ^2+4\lambda
^2\Omega ^2}},\tan\delta =\frac{2\lambda }{\Omega ^2-\omega _0^2}. 
\]

\section{Fluctuation-assisted depinning of the kink}

We will use Langevin approach to describe fluctuations of the kink in the
potential of the pinning centre, i.e. fluctuation force $\tilde f(t)$ needs
to be introduced into the right hand side of the equation \cite{Kl}: 
\begin{equation}
\ddot X+2\lambda \dot X+\omega _0^2\left( X-X_0\right) =\pm \frac{f_0}M\cos
\Omega t+\frac{\tilde f(t)}M.  \label{eq20}
\end{equation}
We consider the fluctuations as being white noise, i.e. the average value of
the fluctuation force equals to zero, and the time correlation is determined
by the delta-function: 
\begin{equation}
\begin{array}{l}
\langle \tilde f(t)\rangle =0, \\ 
\langle \tilde f(t)\tilde f(t^{\prime })\rangle =2D\delta (t-t^{\prime }),
\end{array}
\label{eq21}
\end{equation}
where $D$ is the intensity of the Langevin source (it also represents the
coefficient of diffusion in velocity space). From now on, we will not
consider quantum fluctuations (i.e. we assume that the process temperature
is higher than the Debye temperature of the crystal $T>>\theta _D)$; in this
case $D=\lambda k_BT/M$, where $k_B$ is the Boltzmann constant.

Before proceeding to solution of the Langevin equation, let's determine the
value of the critical shift of the kink centre, which corresponds to the
kink depinning. Obviously, a parabolic potential corresponds to the
linearized equation (\ref{eq17}). However, we have not yet determined the
radius of that potential corresponding to the critical shift. In order to
determine that radius, let's refer to first equation (\ref{eq16a}). We can
determine the local potential of the interaction of kink and pinning centre $%
U(X)$ by integrating over $X$ the last term of this equation (characterizing
the strength of interaction). Taking into account that this potential should
become zero when $X\to \infty $, we get 
\begin{equation}
U(X)=-\frac \mu \pi \ sech{^2}\left( \frac{\pi X}w\right) .  \label{eq22}
\end{equation}
Unfortunately, the potential (\ref{eq22}) does not have the natural cutoff
radius. Therefore, let's assume that some small value of the interaction
potential $U=-U_0$ corresponds to the critical shift value $X_{cr}$. The
kink can be considered as detached from the pinning centre if its energy
state is separated from the zero level (corresponding to the free kink) by
the energy $U_0\sim k_BT$: in this case any single-quantum thermal
fluctuation will lead to final exit from the potential well. Thus, critical
shift can be determined by the following expression: 
\begin{equation}
X_{cr}=\frac w\pi \ arcosh\left[ \left( \frac \mu {\pi k_BT}\right) ^{\frac
12}\right] .  \label{eq23}
\end{equation}
Now we can proceed to calculation of the probability of the
fluctuation-assisted shift of the kink beyond $X_{cr}$ value (\ref{eq23}).
The solution of equation (\ref{eq20}) is the sum of the deterministic shift (%
\ref{eq17}) and shift $\xi \left( t\right) $ determined by fluctuations: 
\begin{equation}
X=X_0+s\left( t\right) +\xi \left( t\right) .  \label{eq24}
\end{equation}
By using the theory of random process overshoots \cite{Ri}, we can determine
the depinning probability of the kink affected by the periodic force as
equal to the average (during oscillation period) speed of positive
overshoots of the sum of stochastic and harmonic shifts beyond the value $%
C=X_{cr}-X_0$:

\begin{equation}
\nu _\Omega =\frac \Omega {2\pi }\int\limits_0^{2\pi /\Omega
}dt\int\limits_0^\infty dv\,vW\left( C-s\left( t\right) ;v-\dot s\left(
t\right) \right)  \label{eq25}
\end{equation}
where $W$ is the joint probability density for the coordinate and velocity
of the random parameter. Since we analyze influence of the white noise $%
\tilde f(t)$ on the system described by linear equation (\ref{eq20}), then
the response (\ref{eq24}) should be a normal process described by Gauss
function: 
\begin{equation}
W\left( X,v\right) =\left( {2\pi \sigma \sigma _1}\right) ^{-1}\exp \left[ -%
\frac{\left( X-s\left( t\right) \right) ^2}{2\sigma ^2}\right] \exp \left[ -%
\frac{\left( v-\dot s\left( t\right) \right) ^2}{2\sigma _1^2}\right] ,
\label{eq26}
\end{equation}
where $\sigma $ and $\sigma _1$ are the dispersions of the coordinate and
the velocity of the fluctuating parameter respectively, which are determined
through the correlation function of the process.

The spectral density of the process $S\left( \omega \right) =\left\langle X{%
^2\left( \omega \right) }\right\rangle $ is determined from equations (\ref
{eq20}) and (\ref{eq21}): 
\[
S\left( \omega \right) =\frac D{\left( {\omega ^2-\omega _0^2}\right)
^2+4\lambda ^2\omega ^2}, 
\]
and for the correlation function $K\left( \tau \right) =\frac 1{2\pi
}\int\limits_{-\infty }^\infty S\left( \omega \right) e^{-i\omega \tau
}d\omega $ we find: 
\begin{equation}
K\left( \tau \right) =\frac{\pi k_BT}{2M\omega _0^2}\exp \left( -{\lambda
\left| \tau \right| }\right) \,\left( {\cos \sqrt{\omega _0^2-\lambda ^2}%
\tau +\frac \lambda {\sqrt{\omega _0^2-\lambda ^2}}\sin \sqrt{\omega
_0^2-\lambda ^2}\left| \tau \right| }\right) .  \label{eq27}
\end{equation}
Dispersions of the coordinate and the velocity are determined by $\sigma
=K\left( {\tau =0}\right) $, $\sigma _1={K}^{\prime \prime }\left( {\tau =0}%
\right) $: 
\begin{equation}
\sigma ^2=\frac{\pi k_BT}{2M\omega _0^2},\quad \sigma _1^2=\frac{\pi k_BT}{2M%
}.  \label{eq28}
\end{equation}
By using expressions (\ref{eq26}) and (\ref{eq28}), we can now perform
integration over $v$ in (\ref{eq25}). By denoting by $\Psi $ the phase of
deterministic shift $\Psi =\Omega t+\delta $ and considering that integrand
is periodic as function of $\Psi $ with the period $2\pi $, we will get 
\begin{equation}
\nu _\Omega =\frac 1q\sqrt{\frac 2\pi }\int\limits_0^\pi {\Phi }^{\prime
}\left( c-p\cos \Psi \right) {\ }_1F_1\left( -\frac 12;\frac 12;-\frac
12\left( pq\sin \Psi \right) ^2\right) d\Psi ,  \label{eq29}
\end{equation}
where ${\Phi }^{\prime }\left( z\right) =\frac 1{\sqrt{2\pi }}\exp \left(
-z^2/2\right) $ is the derivative of the probability integral, $_1F_1\left(
a;b;z\right) $ is the degenerate hypergeometric function; $p=B/\sigma $, $%
q=\Omega {/\omega _0}$, and $c=C{/}\sigma $ is the relative level.

In order to explore the dependence of the kink depinning on the stationary
force, we discuss first the case when \textit{ac} force is absent. In the
latter case, for the probability of the kink depinning the following
expression can be easily obtained: 
\begin{equation}
\nu _0=\frac{\omega _0}{2\pi }\exp \left( -\frac{c^2}2\right) .  \label{eq30}
\end{equation}
Dependence of $\nu _0$ on constant force $\Gamma $ is determined through
dependence of the frequency $\omega _0$ and relative level $c=\sqrt{\frac{%
2M\omega _0^2}{\pi k_BT}}\left( X_{cr}-X_0\right) $ on constant shift $X_0$
(while the dependence of $X_0$ on $\Gamma $ is determined by the
transcendental equation (\ref{eq16})). The analytical function $\nu _0\left(
\Gamma \right) $ can be obtained in case when constant force $\Gamma $ is
significantly smaller than the critical value of $\Gamma _c$ ($\Gamma
<<\Gamma _c$). By denoting $y\equiv \tanh \left( \pi X_0/w\right) $, from
equation (\ref{eq16}) we can get the cubical equation: 
\begin{equation}
y^3-y+\frac{2\Gamma }{3\sqrt{3}\Gamma _c}=0.  \label{eq31}
\end{equation}
In the case $\Gamma <<\Gamma _c$ it has the following solution 
\[
y=-\frac 2{\sqrt{3}}\sin \left( {\frac \Gamma {3\Gamma _c}}\right) \quad . 
\]
By limiting to the first order of value $\Gamma /{\Gamma _c}$, we will get
the following dependence of $X_0$ on the force: 
\[
X_0=-\frac w\pi \ artanh\left( {\frac 2{3\sqrt{3}}\frac \Gamma {\Gamma _c}}%
\right) . 
\]
By inserting this expression into (\ref{eq18}), we will find the
relationship between the eigenfrequency of the system and \textit{dc} force: 
\begin{equation}
\omega _0^2=\frac{2\pi \mu }{w^2M}\left( {1-\frac 13\left( {\frac{4\Gamma }{%
3\Gamma _c}}\right) ^2}\right) .  \label{eq32}
\end{equation}
Finally, using equation (\ref{eq32}) the depinning probability is obtained: 
\begin{eqnarray}
\nu _0=\left( \frac \mu {2\pi w^2M}\right) ^{1/2}\left( {1-\frac 13\left( {\ 
\frac{4\Gamma }{3\Gamma _c}}\right) ^2}\right) ^{1/2}  \nonumber \\
\exp \left[ {-}\frac{ 2\mu }{\pi ^2k_BT}{\left( {1-\frac 13\left( {\frac{%
4\Gamma }{3\Gamma _c}} \right) ^2}\right) }\left( \ arcosh\left( \frac \mu
{\pi k_BT}\right) ^{1/2}+\frac 2{3\sqrt{3}}\frac \Gamma {\Gamma _c}\right)
^2\right] .  \label{eq33}
\end{eqnarray}
The dependence of (\ref{eq33}) is presented on Fig.2. Note that for a value
of \textit{dc} force $\tilde \Gamma $ ($\tilde \Gamma <\Gamma _c)$ which is
determined by the equation 
\[
\frac{w\tilde \Gamma }{2\mu }-\ sech^2\left( \frac{\pi X_{cr}}w\right) \tanh
\left( \frac{\pi X_{cr}}w\right) =0, 
\]
the level $C=X_{cr}-X_0$ becomes zero. Then the depinning probability during
the period ${2\pi }/{\omega _0}$ approaches unity, as it is shown by dashed
curve on the graph.

The dependence of probability (\ref{eq29}) on the amplitude of \textit{ac}
force $B=p\sigma $ can also be obtained more explicitly in several limiting
cases. Note that the condition of incomplete equilibrium $q=\Omega {/\omega
_0}<<1$ should be fulfilled for the quasi-stationary processes under
discussion \cite{LL}. Therefore, depending on the force amplitude, two
limiting cases can be realized: $pq<<1$ and $pq>>1$.

a) The small amplitude of the \textit{ac} force, $pq<<1$. In this case we
can derive from equation (\ref{eq29}) an expression linking $\nu _\Omega $
with the probability of detachment in absence of \textit{ac} force $\nu _0$: 
\[
\nu _\Omega =\frac{\nu _0}\pi \int\limits_0^\pi {\exp }\left( cp\cos \Psi -%
\frac{p^2}2\cos ^2\Psi \right) d\Psi 
\]
The dependence of $\nu _\Omega \left( p\right) $ for various values of the
relative level $c$ is depicted on Fig.3. In the case $c<1$ , $\nu _\Omega $
is a monotonous decreasing function relative to signal $p$, i.e. if the
level of $C$ is less than the amplitude of noise $\sigma $, then presence of
external periodic force reduces the depinning probability.

If $c>1$, $\nu _\Omega \left( p\right) $ increases reaching maximum when $%
p=p_m\left( c\right) $: $p_m\cong 2\sqrt{1-c^{-2}}$ for $pc<1$, and $%
p_m\approx c$ for $pc>1$. Thus, when interaction strength $\mu $ ensures
high values of critical shift $X_{cr}$ (and respectively, the level $C$
exceeds the noise amplitude) and $p\approx p_m$, the probability of
detachment can significantly increase. When signal continues to grow ($%
p>p_m) $, the probability reduces as $\nu _\Omega \sim p^{-1}$.

b) The limit of large amplitudes, $pq>>1$. In this case the probable case is 
$p>c$. As it follows from (\ref{eq25}) and (\ref{eq26}), $\nu _\Omega \to
\Omega {/2\pi }$, i.e. the number of overshoots during the period approaches
unity.

Note that the dependence of kink depinning probability on amplitude of 
\textit{ac} force is analogous to the case of dislocation line (string)
detachment from the oscillating stopper \cite{Va}. The latter is true for
dislocation motions in crystals with low Peierls relief (mostly in metals).

\section{Conclusions}

Perturbation approach allows obtaining linear equation of motion for kink in
pinning potential. This equation was used for exploring fluctuation-assisted
depinning of the kink affected by \textit{dc} and \textit{ac} forces. The
obtained behavior of depinning probability on \textit{ac} force amplitude is
rather intriguing. The fact that this force can result either in decrease or
increase of the probability (depending on amplitude of the force and
strength of the stopper) means that recombination enhancement, modeled in
our work by the harmonic \textit{ac} force, can cause both softening and
hardening of the crystal. It should be noted that both these effects were
observed in ionic crystals with low Peierls potential, where dislocation
segments move (under sufficiently high load) without kink formation \cite
{Suz}. However, in crystals with covalent and mixed covalent-ionic bonds,
where the Peierls mechanism of dislocation motion is prevalent, the
softening is more common phenomena (though the hardening was observed too,
for example, in GaAs crystals during investigation of photoplastic effect 
\cite{Md}). Our analysis shows, that hardening of Peierls-type crystal due
to \textit{ac} excitation can occur in the case when the depinning of
migrating kinks is the bottleneck of the process.

It should be noted also that by introducing the \textit{ac} force we
simulated only the oscillations of the kink, while the pinning centre was
assumed to be immobile. Obviously, it is quite possible that the stopper can
also be influenced by \textit{ac} excitation, particularly when this center
is involved in thermal recombination process and, consequently, the heat
release stimulates its oscillations. This can essentially affect the
depinning probability \cite{Zak}. Extension of our study to the case when
both kink and stopper are subjected to \textit{ac} forces can be the issue
of further development.

\end{document}